%% file: FedGAT.tex
\documentclass[journal,twoside,web]{ieeecolor}

\usepackage{tmi2}
\usepackage{cite}
\usepackage{framed,multirow}
\usepackage{comment}

\usepackage{amsmath,amssymb,amsfonts,mathbbol}
\usepackage{latexsym}
\usepackage{soul}
\usepackage{algorithmic}
\usepackage{graphicx}
\usepackage{textcomp}

\usepackage{adjustbox}
\usepackage{float}
\usepackage{url}
\usepackage{xcolor}

\usepackage{stfloats}
\usepackage{longtable}
\usepackage{multirow}
\usepackage[linesnumbered,ruled,vlined]{algorithm2e}
\usepackage[font=small,justification=justified,belowskip=2pt,aboveskip=2pt]{caption}

\renewcommand{\arraystretch}{1.4}
\usepackage{hyperref}
\hypersetup{
    colorlinks=true,
    linkcolor=blue,
    filecolor=magenta,      
    urlcolor=cyan,
    pdftitle={FedGAT},
}

\SetKwInput{KwInput}{Input}                
\SetKwInput{KwOutput}{Output}              

\floatname{algorithm}{Algorithm}

\begin{document}
\title{Generative Autoregressive Transformers for Model-Agnostic Federated MRI Reconstruction}
\author{Valiyeh A. Nezhad, Gokberk Elmas, Bilal Kabas, Fuat Arslan, Emine U. Saritas, and Tolga \c{C}ukur \vspace{-1.35cm}
\\
\thanks{\\
This study was supported in part by a TUBA GEBIP 2015 fellowship, and a BAGEP 2017 fellowship (Corresponding author: Tolga Çukur).}
\thanks{V.A. Nezhad, G. Elmas, B. Kabas, F. Arslan, E.U. Saritas, and T. Çukur are with the Department of Electrical and Electronics Engineering, and the National Magnetic Resonance Research Center, Bilkent University, Ankara, Turkey (e-mail: \{valiyeh.ansarian@, gokberk@ee., bilal.kabas@, fuat.arslan@, saritas@ee., cukur@ee.\}bilkent.edu.tr).}
}

\maketitle

\begin{abstract}
While learning-based models hold great promise for MRI reconstruction, single-site models trained on limited local datasets often show poor generalization. This has motivated collaborative training across institutions via federated learning (FL)—a privacy-preserving framework that aggregates model updates instead of sharing raw data. Conventional FL requires architectural homogeneity, restricting sites from using models tailored to their resources or needs. To address this limitation, we propose FedGAT, a model-agnostic FL technique that first collaboratively trains a global generative prior for MR images, adapted from a natural image foundation model composed of a variational autoencoder (VAE) and a transformer that generates images via spatial-scale autoregression. We fine-tune the transformer module after injecting it with a lightweight site-specific prompting mechanism, keeping the VAE frozen, to efficiently adapt the model to multi-site MRI data. In a second tier, each site independently trains its preferred reconstruction model by augmenting local data with synthetic MRI data from other sites, generated by site-prompting the tuned prior. This decentralized augmentation improves generalization while preserving privacy. Experiments on multi-institutional datasets show that FedGAT outperforms state-of-the-art FL baselines in both within- and cross-site reconstruction performance under model-heterogeneous settings.
\vspace{-0.10cm}
\end{abstract}

\begin{IEEEkeywords}
MRI reconstruction, federated learning, generative, autoregressive transformer. \vspace{-0.35cm}
\end{IEEEkeywords}

\bstctlcite{IEEEexample:BSTcontrol}

\section{Introduction}
Magnetic Resonance Imaging (MRI) is an indispensable diagnostic tool due to its superior soft-tissue contrast. However, its inherently long scan times can impede clinical workflow and reduce patient comfort \cite{Lustig2007}. To accelerate acquisitions, a common strategy is to reconstruct images from undersampled k-space measurements \cite{Zhao2015}. Recent learning-based methods have been effective in recovering high-quality images from undersampled acquisitions \cite{Hammernik2017, Kwon2017, Dar2017, Zhu2018, raki}. Yet, models trained exclusively on local datasets often fail to generalize well, especially when the local data distribution shows limited diversity in tissue characteristics and acquisition protocols \cite{Schlemper2017, MoDl, Quan2018c, Mardani2019b, rgan, PatelRecon}. Robust, generalizable reconstruction models require training on diverse multi-site datasets—an objective that might be hindered by privacy concerns, institutional restrictions, and economic constraints \cite{LiangSPM, chen2022review}. These challenges underscore the need for collaborative learning frameworks that enable multi-site model development while preserving data privacy.

Several strategies might be considered to enhance cross-site generalization in reconstruction models. One option is centralized training, where institutions share raw MRI data in a common repository \cite{kaissis2020secure}. Although straightforward in principle, this approach is only occasionally feasible because of strict privacy regulations and the difficulty of maintaining large centralized repositories \cite{FedGAN,Sheller2019}. Synthetic data sharing offers an alternative approach, where sites generate and exchange synthetic MRI data that mimic their local data distributions \cite{chang2023mining}. However, transferring synthetic MR data can incur significant communication and storage overhead, especially as the number of samples per site and participating sites increases \cite{kang2025efficient}. Another option involves each site independently pretraining its reconstruction model on a public dataset, followed by local fine-tuning on the target dataset \cite{Dar2017}. While this strategy can reduce the need for data sharing, it may be limited by the representational scope of public datasets, which do not always reflect the diversity of anatomies or imaging protocols encountered in clinical settings.

Among collaborative strategies, federated learning (FL) has emerged as a promising approach for privacy-preserving, scalable model development across institutions \cite{WenqiLi2019, Li2020}. In FL, sites keep their imaging data local and contribute to collaborative training by transmitting model updates rather than raw data. Recent studies have reported improved generalization across institutions in MRI reconstruction by adopting the conventional FL framework, in which locally trained reconstruction models are iteratively aggregated into a globally shared reconstruction model \cite{guo2021, elmas2022federated, feng2023tmi, wang2024fed, levac2023bio}. Nonetheless, a notable limitation of conventional FL is the requirement that all participating sites use the same network architecture to allow aggregation of model weights across sites.  

In practice, imaging sites can differ in clinical priorities, computational resources, and data distributions, leading to varying preferences in network architectures for MRI reconstruction. Some facilities may favor convolutional models for their spatial precision and computational efficiency \cite{guo2021, elmas2022federated}; others may opt for transformers to better capture long-range dependencies for context-aware processing \cite{feng2023cvpr}, while physics-driven unrolled networks may be preferred for their robustness to variations in imaging operators such as sampling densities and coil arrays. \cite{levac2023bio, wang2024fed}. Enforcing a shared backbone across all sites in conventional FL can therefore discourage participation and constrain performance. Meanwhile, conducting separate FL sessions for each architecture of interest requires retraining models from scratch, introducing substantial computational overhead for resource-limited institutions. Although recent FL methods have explored model personalization strategies to address inter-site data heterogeneity through techniques such as partial aggregation \cite{feng2023tmi}, test-time adaptation \cite{elmas2022federated}, or feature map normalization \cite{dalmaz2024one}, these techniques still operate under a fixed global architecture. Existing FL methods for MRI reconstruction thus remain incompatible with fully model-heterogeneous FL settings.

\begin{figure*}[t]
\centering\includegraphics[width=0.87\textwidth]{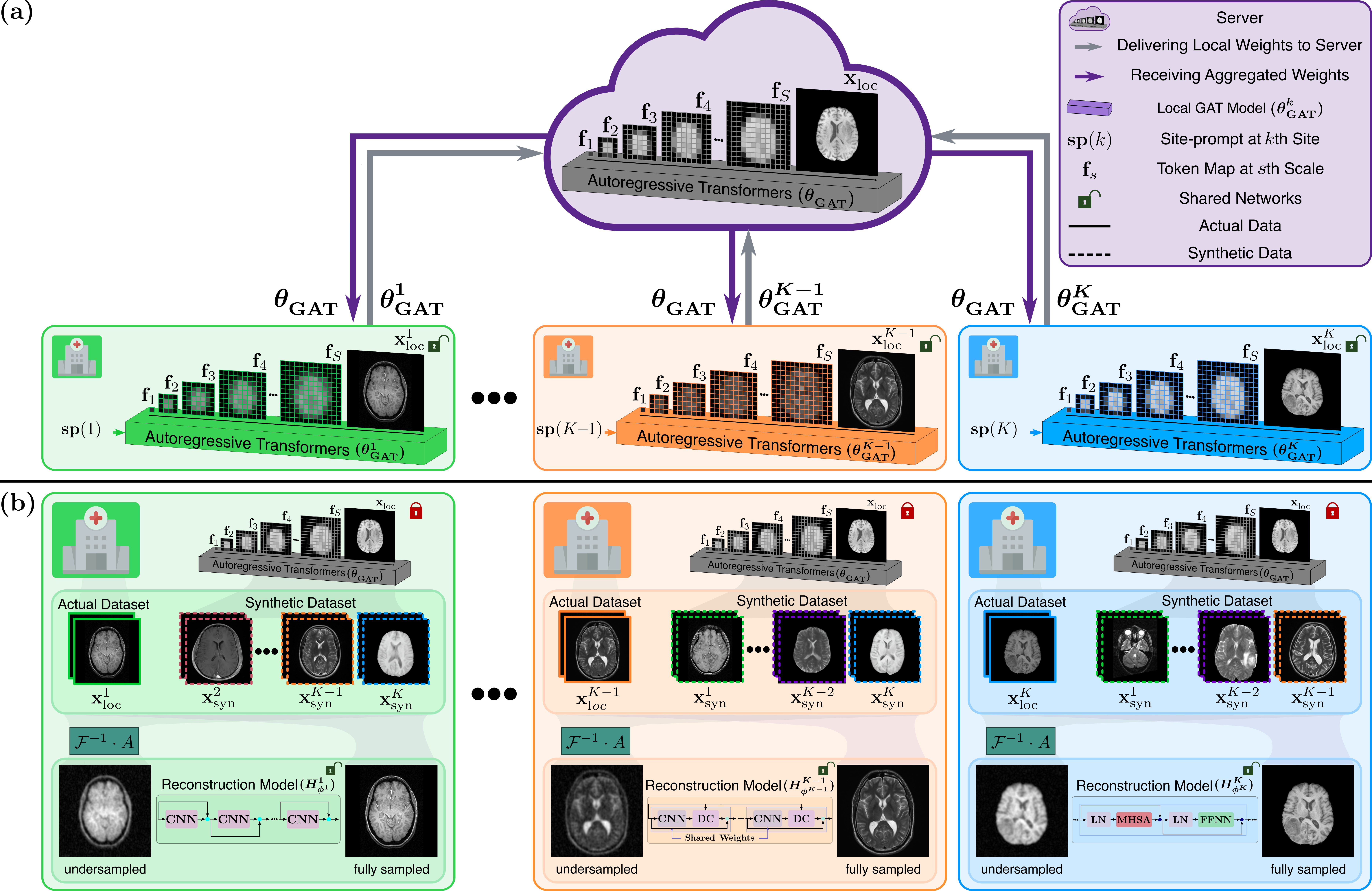}
\captionsetup{justification=justified, width=\textwidth}
\caption{FedGAT adopts a two-tier framework to enable collaborative training of heterogeneous MRI reconstruction models. \textbf{(a)} The first tier conducts decentralized training of a global prior ${\theta_{\text{GAT}}}$, a generative autoregressive transformer that models the distribution of multi-site MR images via autoregressive prediction across increasing spatial scales, guided by a site prompt $\texttt{sp}$ to retain site-specific attributes. \textbf{(b)} The second tier carries out local training of site-specific reconstruction models $H^k_{{\phi}^k}$ $(k$: site index) on hybrid datasets that combine local MRI data with GAT-generated synthetic images emulating remaining sites, thereby promoting generalization while preserving data privacy.}\vspace{-0.25cm}
\label{fig:FedGAT_gen}
\end{figure*}

Here, we propose a novel technique for model-heterogeneous collaborations, \textit{FedGAT}, the first model-agnostic FL method for MRI reconstruction to our knowledge (see a preliminary version in \cite{fedgat_ismrm}). In the broader FL literature, distillation-based approaches have been proposed for model-heterogeneous settings—transferring knowledge between teacher and student copies of local and global models—but they incur high computational cost and often achieve limited cross-site model alignment. As an alternative, generative methods based on adversarial or diffusion priors have been considered, as they can lower compute demand by decoupling knowledge transfer regarding data distributions from model training \cite{elmas2022federated}. However, their use in FL can be limited by unstable training (adversarial), high inference costs (diffusion), and difficulty in fine-grained control over site-specific outputs \cite{MONAI, guo2024maisimedicalaisynthetic}. To address these challenges, we construct a generative autoregressive transformer (GAT) prior atop a foundation model originally designed for natural image synthesis, comprising a variational autoencoder (VAE) and an autoregressive transformer \cite{VAR}. Training such priors from scratch on medical data is often impractical due to data scarcity and computational demands. Therefore, we introduce a lightweight, site-specific prompting mechanism into the transformer module and efficiently fine-tune only this module while keeping the VAE frozen. The transformer weights are collaboratively updated across sites, enabling distributed adaptation without sharing raw data. Afterwards, each site expands its local MRI dataset by incorporating multi-site synthetic images generated by the GAT prior, and trains its reconstruction model using this augmented dataset. 

The GAT prior synthesizes images via spatial-scale autoregression, balancing image quality, training stability, and computational and storage efficiency. Site-specific prompting also enables explicit control over the output distribution. These features make GAT well suited for FL in which fidelity, stability, and efficiency are critical \cite{kang2025efficient}. Since the synthetic data can be reused to train any architecture at individual sites independently of others, FedGAT substantially reduces the computational overhead associated with model training. It further enhances scalability, as new sites can readily develop their own models by augmenting local datasets with synthetic samples from the shared GAT prior, without requiring global retraining. Experiments on multi-institutional MRI datasets show that FedGAT enables successful collaboration across heterogeneous model architectures and outperforms existing model-agnostic FL methods in both within- and cross-site reconstruction performance. Code is publicly available at: {\small \url{https://github.com/icon-lab/FedGAT}}.

\vspace{0.1cm}
\subsubsection*{\textbf{Contributions}} 
\begin{itemize} 
\item To our knowledge, we introduce the first model-agnostic FL framework for MRI reconstruction, enabling model-heterogeneous collaboration across sites.
\item We propose to decouple federated training of a global generative prior from local building of site-specific reconstruction models to support model heterogeneity.
\item We develop a federated GAT prior with spatial-scale autoregression and site-prompting to yield high-fidelity, site-consistent synthetic MRI data.
\item We design a hybrid procedure that trains site-specific reconstruction models on mixed local-synthetic datasets to balance performance with generalization.
\end{itemize}

\section{Theory}

\subsection{Conventional FL for MRI Reconstruction}
Image and k-space domains in MRI are linked through the imaging operator $\mathcal{A} = \mathcal{M} \mathcal{F} \mathcal{C}$, where $\mathcal{M}$ is the sampling pattern, $\mathcal{F}$ is Fourier transform, and $\mathcal{C}$ are coil sensitivities:
\begin{equation}
\label{eq:AXY}
\mathcal{A} \mathbf{x} = \mathbf{y},
\end{equation}
where $\mathbf{x}$ is the image and $\mathbf{y}$ is acquired data. For undersampled acquisitions, solving Eq. \ref{eq:AXY} requires the incorporation of a regularization term $\mathcal{R}(\cdot)$ \cite{Lustig2007}:
\begin{equation}
\label{eq:regularized_recon}
\widehat{\mathbf{x}} = \underset{\mathbf{x}}{\operatorname{argmin}} || \mathbf{y} - \mathcal{A} \mathbf{x} ||_{2}^{2} + \mathcal{R}(\mathbf{x}).
\end{equation}
Recent learning-based methods solve Eq. \ref{eq:regularized_recon} using network projections, typically in the image domain: $\widehat{\mathbf{x}} = H_{\phi}(\mathcal{F}^{-1} \mathbf{y}, \mathcal{A})$, where $H_{\phi}$ takes a zero-filled reconstruction as input \cite{Hammernik2017}.

In conventional FL for MRI reconstruction, a model is collaboratively trained over $N_c$ communication rounds between a central server and $K$ sites \cite{WenqiLi2019}. The server maintains a global model $H_{\phi}$ with weights $\phi$, and each site holds a local copy $H_{\phi^k}$ for $k = 1, ..., K$. At round $c$, sites initialize with the current global weights: $\phi^k \gets \phi(c-1)$, then train $H_{\phi^k}$ on local data $\mathcal{D}_{\text{loc}}^k$ to minimize:
\begin{equation}
\label{eq:fed_dnn_training}
\mathcal{L}^{k}_{\text{rec}} = \mathbb{E}{(\mathbf{x}^{k}, \mathbf{y}^{k}) \sim \mathcal{D}_{\text{loc}}^k} \left[ || \mathbf{x}^{k} - H_{\phi^k}(\mathcal{F}^{-1}\mathbf{y}^{k}, \mathcal{A}^{k}) ||_2^2 \right],
\end{equation}
where $\mathbf{x}^{k}$ is the fully-sampled reference. Afterwards, models are sent to the server and aggregated via averaging \cite{FedAvg}:
\begin{equation}
\phi(c) = \sum_{k=1}^{K} \alpha_k \phi^k,
\end{equation}
where $\alpha_k = \frac{N^k}{\sum_{j=1}^{K} N^j}$ and $N^k$ is the number of training samples at site $k$. After $N_c$ rounds, the final global model $H_{\phi^*}$ with $\phi^* := \phi(N_c)$ is deployed. For test data at site $k$:
\begin{equation}
\widehat{\mathbf{x}}\hspace{0.1em}^{k} = H_{\phi^*}(\mathcal{F}^{-1}\mathbf{y}^{k}, \mathcal{A}^{k}),
\end{equation}
where $\widehat{\mathbf{x}}\hspace{0.1em}^{k}$ is image reconstructed from $\mathbf{y}^{k}$.

\subsection{Federated Generative Autoregressive Transformers}
FedGAT is a model-agnostic FL technique that supports collaborative training of heterogeneous architectures across sites via a two-tier strategy. In the first tier, a federated GAT prior is trained to capture the distribution of multi-site MR images (Fig. \ref{fig:FedGAT_gen}\textbf{a}). In the second tier, site-specific reconstruction models are locally trained (Fig. \ref{fig:FedGAT_gen}\textbf{b}). Each site starts training a reconstruction model of preferred architecture on its local dataset, then continues training on a hybrid dataset of local data and synthetic data from other sites generated via the prior. The architecture and training procedures are detailed below.

\vspace{0.25cm}
\subsubsection{\underline{Architecture of the Site-Prompted GAT Prior}}
\label{sec:architecture}
FedGAT employs a global generative prior equipped with a site-prompt to model the distribution of multi-site MR images, enabling selective generation of synthetic data from individual sites (Fig. \ref{fig:GAT_arch}). Our approach builds upon a foundation generative model for natural images—a VAE combined with an autoregressive transformer (VAR) \cite{VAR}—which has demonstrated strong contextual modeling capabilities in natural image synthesis. Motivated by this success and recent advances in transformers for MR image formation \cite{MTrans}, we adapt this foundation to MRI by introducing a novel site-prompting mechanism that conditions the autoregressive transformer on site identity, enabling controlled synthesis of diverse MR images across sites.

Transformer architectures process images as token sequences with quadratic complexity in sequence length, making pixel-level application computationally expensive and patch-level tokens less spatially precise \cite{vaswani2021scaling}. To balance this, VAR synthesizes feature maps in a compact latent space, with image-latent mappings handled by the VAE encoder-decoder \cite{VQVAE}. Our GAT prior retains this compound architecture but focuses fine-tuning on the transformer module with learnable site-specific prompts, while keeping the VAE frozen, to efficiently adapt the foundation model to multi-site MRI data.

\textit{\textbf{VAE Encoder Module:}} Given a coil-combined complex MR image $\mathbf{x} \in \mathbb{R}^{H \times W \times 2}$, where real and imaginary parts form two channels, the encoder maps $\mathbf{x}$ to discrete token maps $\mathbf{f}_1, \dots, \mathbf{f}_S$ across $S$ spatial scales. It first computes a continuous latent $\mathbf{z} \in \mathbb{R}^{h_S \times w_S \times c}$ at the highest scale ($S$) using residual convolutional blocks: $\mathbf{z} = \mathrm{ResConv}(\mathbf{x})$. Discrete token maps at scale $s$ are then obtained by downsampling $\mathbf{z}$ to $h_s \times w_s$, followed by quantization using a learnable codebook $\mathbf{B} \in \mathbb{R}^{V \times c}$ of vocabulary size $V$, defining a discrete latent space with $V$ categories.

To reduce redundancy and information loss across scales, we adopt a hierarchical token extraction procedure progressing from the lowest ($s = 1$) to the highest ($s = S$) spatial scale \cite{VQGAN}. A residual representation $\mathbf{r}_s \in \mathbb{R}^{h_S \times w_S \times c}$ is maintained, initialized as $\mathbf{r}_1 = \mathbf{z}$. At each scale $s$, $\mathbf{f}_s$ is obtained by quantizing the downsampled residual:
\begin{equation}
\mathbf{f}_s = \underset{v \in \{1, \dots, V\}}{\mathrm{argmin}} \, \lVert \mathbf{B}(v\,,\,:) - \text{Down}_s(\mathbf{r}_{s}) \rVert^2_2,
\label{eq:quant}
\end{equation}
where $\mathbf{f}_s \in [V]^{h_s \times w_s}$, and $\text{Down}_s$ is downsampling interpolation. Quantization is performed via nearest-neighbor lookup in $\mathbf{B}$ using Euclidean distance. Following \cite{VQVAE}, $\mathbf{B}$ is initialized uniformly at random, which—along with the objective in Eq. \ref{eq:quant}—yields a uniform prior over token maps $\mathbf{f}_s$.

Codebook vectors corresponding to $\mathbf{f}_s$ are then upsampled to the $S$th scale and used to update the residual:
\begin{equation}
\mathbf{r}_{s+1} = \mathbf{r}_{s} - \text{Conv}(\text{Up}_S(\text{Lookup}(\mathbf{B},\mathbf{f}_s))),
\end{equation}
where $\text{Lookup}$ retrieves codebook vectors, $\text{Up}_S$ is upsampling interpolation, and $\text{Conv}$ is a convolutional layer. The encoder outputs token maps aggregated across scales:
\begin{equation}
\{\mathbf{f}_1, \mathbf{f}_2, \dots, \mathbf{f}_S\} = \mathrm{Enc}(\mathbf{x}),
\end{equation}
yielding a multi-scale representation of image features.

\begin{figure*}[!t]
       \vspace{-0.25cm}
        \centering
        \captionsetup{justification=justified, singlelinecheck=false}
        \includegraphics[width=\textwidth]{figures/figure02.png}
        \caption{Architecture of the proposed site-prompted GAT prior. \textbf{(a)} The GAT prior embodies a variational autoencoder (VAE), whose encoder maps an input MR image onto a set of discrete token maps $\mathbf{f}_1, \mathbf{f}_2, \dots, \mathbf{f}_S$ across $S$ spatial scales and the decoder reconstructs the image from these token maps. \textbf{(b)} The transformer module establishes an autoregressive prior over the multi-scale token maps, predicting each higher-scale map $\mathbf{f}_s$ conditioned on preceding maps $\mathbf{f}_{<s} := \{\mathbf{f}_1, ..., \mathbf{f}_{s-1}\}$. To retain site-specific features in synthetic MR images, a site prompt $\mathbf{sp}(k)$ is derived from a one-hot site index via a gated MLP (gMLP), and used to initialize the site token $\mathbf{st}$ in the transformer.}  
        \vspace{-0.25cm}
        \label{fig:GAT_arch}
\end{figure*}

\textbf{\textit{VAE Decoder Module:}} The VAE decoder reconstructs the MR image $\mathbf{x}$ from the discrete multi-scale token maps $\{\mathbf{f}_1, \mathbf{f}_2, \dots, \mathbf{f}_S\}$. Mirroring the encoder’s extraction, it follows a hierarchical reconstruction from low to high spatial scales \cite{VQGAN}. A residual representation is initialized as $\hat{\mathbf{r}}_{0}=0 \in \mathbb{R}^{h_S \times w_S \times c}$. At each scale $s$, codebook vectors for $\mathbf{f}_s$ are retrieved, upsampled, and used to update the residual via:
\begin{equation}
\hat{\mathbf{r}}_{s} = \hat{\mathbf{r}}_{s-1} + \text{Conv}(\text{Up}_S(\text{Lookup}(\mathbf{B},\mathbf{f}_s))).
\end{equation}
After all $S$ scales, the final continuous representation $\hat{\mathbf{z}} = \hat{\mathbf{r}}_S \in \mathbb{R}^{h_S \times w_S \times c}$ is projected through residual convolutional blocks to recover the image:
\begin{eqnarray}
\label{eq:vaedec}
\hat{\mathbf{x}} = \text{ResConv}(\hat{\mathbf{r}}_S)=\mathrm{Dec}(\{\mathbf{f}_1, \mathbf{f}_2, \dots, \mathbf{f}_S\}),
\end{eqnarray}
where $\hat{\mathbf{x}} \in \mathbb{R}^{H \times W \times 2}$ is the predicted MR image.

\textit{\textbf{Autoregressive Transformer Module:}} 
While the VAE encoder extracts discrete token maps across spatial scales, it does not explicitly model statistical dependencies among them \cite{VQVAE}. \textit{To capture the joint distribution of $\mathbf{f}_1,..,\mathbf{f}_S$, we propose an autoregressive prior over multi-scale token maps using a transformer module.} Specifically, the transformer predicts tokens $\mathbf{f}_s$ at scale $s$ conditioned on preceding tokens $\mathbf{f}_{<s}:=\{\mathbf{f}_1,...,\mathbf{f}_{s-1}\} \in [V]^{T_{s-1}}$, where $T_{s-1} = \sum_{o=1}^{s-1} h_o w_o$ is the number of tokens at lower scales. To preserve spatial alignment, each $\mathbf{f}_o$ is embedded after upsampling to match the dimensions of scale $s$:
\begin{align}
 \mathbf{e}_o = \mathrm{Lin}(\mathrm{Up}_{o+1}(\mathbf{f}_o)),\mbox{ for }o\in[1 \mbox{ } s-1],
\end{align}
with $\mathbf{e}_o \in \mathbb{R}^{h_{o+1} w_{o+1} \times d}$, $\mathbf{e}_{<s}$$\,:=\,$$\{\mathbf{e}_1,...,\mathbf{e}_{s-1}\} \in \mathbb{R}^{(T_s-1) \times d}$. 

\textit{To align feature maps with site-specific data distributions, we propose a learnable site prompt $\mathbf{sp}(k) \in \mathbb{R}^{1 \times d}$ derived by projecting the one-hot site encoding $\mathbb{1}_{K}(k)$ through a gated multi-layer perceptron (gMLP)}:
\begin{align}
    \mathbf{g} = \mathrm{GELU}&(\mathrm{Lin}(\mathbb{1}_{K}(k))), \quad
    \mathbf{g}_u, \mathbf{g}_v = \mathrm{Split}(\mathbf{g}), \\
    \mathbf{sp}(k) &= \mathrm{Lin}(\mathrm{Lin}(\mathrm{LN}(\mathbf{g}_u)) \odot \mathbf{g}_v),
\end{align}
where GELU is the activation, $\mathbf{g}$ is the hidden output, and $\mathbf{g}_u$, $\mathbf{g}_v$ are channel-wise splits. The resulting site token $\mathbf{st} = \mathbf{sp}(k)$ is pooled with token embeddings and combined with position and scale encodings:
\begin{equation}
\label{eq:at_first}
\mathbf{h}_0 = \Big[ \big[\mathbf{st};\,\, \mathbf{e}_{<s}\cdot \mathbf{E}_{\text{tok}}\big] + \mathbf{E}_{\text{pos}} + \mathbf{E}_{\text{sca}};\,\, \mathbf{m} \Big],
\end{equation}
where $\mathbf{h}_0 \in \mathbb{R}^{T_S\times d}$ is the transformer input, $\mathbf{E}_{\text{tok}} \in \mathbb{R}^{(T_s-1) \times d}$ is the token projection matrix, $\mathbf{E}_{\text{pos}} \in \mathbb{R}^{T_s \times d}$ encodes spatial positions \cite{attention}, and $\mathbf{E}_{\text{sca}} \in \mathbb{R}^{T_s \times d}$ encodes scales. The zero matrix $\mathbf{m}$$\,=\,$$0^{(T_S-T_{s}) \times d}$ masks finer-scale tokens, enforcing causal progression across spatial scales. 

To predict $\mathbf{f}_s$, the hidden representation $\mathbf{h}_0$ is passed through $L$ transformer blocks, each comprising multi-head self-attention (MHSA) and MLP layers interleaved with normalization and residual connections \cite{vaswani2021scaling}. Unlike conventional transformers, we employ adaptive layer normalization (AdaLN) to enable site-specific processing of hidden representations. Specifically, AdaLN modulates the statistics of $\mathbf{h}$ via learnable functions of the site token $\mathbf{st}$:
\begin{equation}
\mathrm{AdaLN}(\mathbf{h}, \mathbf{st}) = \gamma(\mathbf{st}) \cdot \frac{\mathbf{h} - \mu(\mathbf{h})}{\sigma(\mathbf{h})} + \beta(\mathbf{st}),
\end{equation}
where $\mu$ and $\sigma$ are the mean and standard deviation of $\mathbf{h}$, and $\beta$, $\gamma \in \mathbb{R}$ are learnable bias and gain parameters conditioned on $\mathbf{st}$. The projection through the $\ell$th transformer block ($\ell \in [1 \mbox{ } L]$) is defined as:
\begin{eqnarray}
\mathbf{h}^\prime_{\ell} &=& \mathbf{h}_{\ell-1} + \mathrm{MHSA}(\mathrm{AdaLN}(\mathbf{h}_{\ell-1}, \mathbf{st})),\\
\mathbf{h}_{\ell}&=& \mathbf{h}^\prime_{\ell} + \mathrm{MLP}(\mathrm{AdaLN}(\mathbf{h}^\prime_{\ell}, \mathbf{st})).
\end{eqnarray}
In MHSA, attention scores between query and key tokens in $\mathbf{h}$ are computed via softmax dot-product similarity \cite{attention}, and used to filter the value tokens:
\begin{equation}
A_{i,j} = \text{Softmax}\left(\frac{\mathbf{q}_i \cdot \mathbf{k}_j^\top}{\sqrt{d}}\right); \quad
\mathrm{SA}(\mathbf{h}) = A \cdot \mathbf{v},
\end{equation}
where $A \in \mathbb{R}^{(1+T_S) \times (1+T_S)}$ is the attention score matrix, $i$, $j$ are token indices, and $\mathbf{q}$, $\mathbf{k}$, $\mathbf{v}$ are unit-norm query, key, and value tokens obtained via learnable linear projections of $\mathbf{h}$. $\mathrm{SA}$ denotes a single self-attention head in MHSA.

After the final transformer block, an adaptive layer norm conditioned on $\mathbf{st}$ and a linear layer are applied to extract the logits $\mathbf{logit} \in \mathbb{R}^{(h_s w_s) \times V}$, enhancing site specificity. The token probabilities $\mathbf{P}_s$ are computed from these logits, and the most probable codebook indices are selected to obtain the predicted token sequence:
\begin{eqnarray}
\mathbf{logit} &=& \mathrm{Lin}(\mathrm{AdaLN}(\mathbf{h}_{L}(T_{s-1}+1:T_{s},:),\mathbf{st}))\\
\mathbf{P}_s &=& \mathrm{Softmax}({\mathbf{logit}}), \label{eq:mp1}\\
\hat{\mathbf{f}}_s(i) &=& \underset{v \in \{1, \dots, V\}}{\mathrm{argmax}}\, \mathbf{P}_s(i,v),\label{eq:mp2}
\end{eqnarray}
where $\hat{\mathbf{f}}_s(i)$ denotes the predicted index for the $i$th token in $\mathbf{f}_s$, and $\mathbf{P}_s(i,v)$ is the predicted probability of the $v$th codebook vector. For brevity, we denote the full autoregressive mapping performed by the transformer as:
\begin{equation}
\hat{\mathbf{f}}_s = \mathrm{Trans}(\mathbf{sp}(k),\,\mathbf{f}_{<s}), \mbox{for } s \in [1\mbox{ }{S}]; \{\mathbf{f}_0\} \in \varnothing. 
\label{eq:at_last}
\end{equation}

\subsubsection{\underline{MR Image Synthesis with the GAT Prior}}
\label{sec:synthesis}
To synthesize multi-site MR images using the GAT prior, we employ the autoregressive transformer and VAE decoder modules. The transformer generates a random set of multi-scale discrete token maps sequentially across increasing spatial scales, following the autoregressive mapping in Eq. \ref{eq:at_last}. For stochastic inference, we replace the maximum-probability sampling in Eqs.~\ref{eq:mp1}–\ref{eq:mp2} with nucleus (top-$p$) sampling. The logits over $V$ codebook entries are sorted in descending order to compute cumulative probabilities $\mathbf{c}(i,:)$. Tokens outside the smallest subset satisfying $\mathbf{c}(i,k)\!\ge\!p$ are masked, and remaining logits are normalized to form the sampling distribution:
\begin{eqnarray}
\mathbf{logit}_{\mathrm{sort}}(i,:) &=& \mathrm{Sort}\big(\mathbf{logit}(i;1\!:\!V),\mathrm{'descend'}\big),\\
\tilde{\mathbf{P}}_s(i,:) &=& \mathrm{Softmax}\big(\mathbf{logit}_{\mathrm{sort}}(i,\mathbf{c}(i,:)\!\le\!p)\big),\\
\hat{\mathbf{f}}_s(i) &\sim& \tilde{\mathbf{P}}_s(i,:),
\end{eqnarray}
where $\tilde{\mathbf{P}}_s(i,:)$ is the renormalized top-$p$ distribution.

Once a set of discrete multi-scale token maps $\{\mathbf{f}_1, \mathbf{f}_2, \dots, \mathbf{f}_S\}$ is generated, it can be projected through the VAE decoder as described in Eq. \ref{eq:vaedec} to obtain a synthetic MR image. For brevity, here we refer to the overall mapping used to synthesize images from site $k$ as follows:
\begin{align}
 \mathbf{x}^k_{\text{syn}} &= \mathrm{GAT}(\mathbf{sp}(k)), \mbox{ such that:} \\
 \mathrm{GAT}(\cdot) &= \mathrm{Dec} \left( \left\{ (\mathbf{f}_1,\dots,\mathbf{f}_S):\, \right. \right. \nonumber \\ 
 &\qquad \quad \left. \left. \mathbf{f}_s = \mathrm{Trans}(\cdot,\mathbf{f}_{<s}) \mbox{ for } s\in[1\mbox{ }S];\right\} \right).\nonumber
\end{align}

\begin{algorithm}[t]\small
\caption{Training of reconstruction models}\label{alg:site_specific_recon}
\small
\KwInput{
    $\mathcal{D}^k_{\text{\text{loc}}}$: local dataset from site $k$; $\text{GAT}$: global prior. \\
    $N_p$: num. of pre-training, $N_f$: num. of follow-up epochs. \\
}
\KwOutput{
   $H^k_{{\phi}^k}$: reconstruction model at site $k$.
}
Initialize model with $\phi^k(0)$ at site $k$.\\
{$\triangleright$ pre-train on local dataset}\\
\For{$e = 1$ \textbf{to} $N_p$}{ 
    $\phi^k(e) \leftarrow \phi^k(e-1) - \eta \nabla_{\phi^k}\mathcal{L}^k_{\text{rec}}(\mathcal{D}^k_{\text{\text{loc}}})$ 
}
{$\triangleright$ generate synthetic datasets}\\
\For{$j = 1$ \textbf{to} $K$}{
    $\mathcal{D}^j_{\text{syn}} = \{ \mathbf{x}^j_{\text{syn}}, \mathcal{A}^k \mathbf{x}^j_{\text{syn}} \}$ where $\mathbf{x}^j_{\text{syn}} = \text{GAT}(\mathbf{sp}(j))$\\
}
{$\triangleright$ follow-up train on local \& synthetic datasets}\\
\For{$e = (1+N_p)$ \textbf{to} $(N_f+N_p)$}{
    $\phi^k(e)$$\leftarrow$$\phi^k(e$$-$$1)$$-$$\eta \nabla_{\phi^k}\mathcal{L}^k_{\text{rec}}([\mathcal{D}^k_{\text{\text{loc}}};\, \mathcal{D}_{\text{syn}}^{\{1,..,K\} \setminus \{k\}}])$ 
}
\end{algorithm}

\subsubsection{\underline{Two-Tier Training Procedure for FedGAT}} 
To support model-heterogeneous settings, FedGAT employs a two-tier training strategy, decoupling global GAT prior learning from site-specific reconstruction model training (Alg. \ref{alg:site_specific_recon}). 

\textbf{\textit{Tier 1 – Training the GAT prior:}} The GAT prior comprises a VAE and an autoregressive transformer (Sec. \ref{sec:architecture}). The VAE encodes input MR images into a compact latent space and decodes them back. Its training uses a composite loss \cite{VQGAN}:
\begin{align}
\label{eq:vaeloss}
\mathcal{L}^{k}_{\text{vae}}(\mathcal{D}_{\text{loc}}^k) = \mathbb{E}_{\mathbf{x}^{k} \sim \mathcal{D}_{\text{loc}}^k}  \left[ \| \mathbf{x}^{k} - \hat{\mathbf{x}}^{k} \|_2^2 + \| \mathbf{z}^{k} - \hat{\mathbf{z}}^{k} \|_2^2 \right. && \nonumber \\
\left. + \lambda_p \mathrm{LPIPS}(\mathbf{x}^{k}, \hat{\mathbf{x}}^{k}) + \lambda_a \mathcal{L}_a(\hat{\mathbf{x}}^k) \right], && 
\end{align}
where $k$ is the site index, and $\mathcal{D}_{\text{loc}}^k$ is the local dataset. The four terms respectively enforce decoding fidelity, latent consistency, perceptual similarity, and adversarial realism. Meanwhile, the transformer learns to autoregressively predict multi-scale discrete token maps, assigned categorically via the VAE codebook. It is trained using a cross-entropy loss:
\begin{align}
\label{eq:transloss}
\mathcal{L}^{k}_{\text{trans}}(\mathcal{D}_{\text{loc}}^k) = - \mathbb{E}_{\mathbf{x}^k \sim \mathcal{D}_{\text{loc}}^k} \left[ \sum_{s=1}^{S} \sum_{i=1}^{h_s w_s} \sum_{v=1}^{V} \mathbb{I}(\mathbf{f}_s^k(i) = v)\right. \nonumber \\ 
\cdot \log \left( \mathbf{P}_s(i, v) \right)  \Bigg],
\end{align}
where $\{\mathbf{f}_1^k, \dots, \mathbf{f}_S^k\} = \mathrm{Enc}(\mathbf{x}^k)$ are the ground-truth token maps, $i$ indexes tokens, and $\mathbb{I}$ indicates if token $\mathbf{f}^{k}_{s}(i)$ matches the $v$th codebook entry. The combined GAT loss is given by:
\begin{align}
\label{eq:gatloss}
\mathcal{L}^{k}_{\text{GAT}}(\mathcal{D}_{\text{loc}}^k) = \mathcal{L}^{k}_{\text{trans}}(\mathcal{D}_{\text{loc}}^k) + \lambda_{\text{vae}} \mathcal{L}^{k}_{\text{vae}}(\mathcal{D}_{\text{loc}}^k),
\end{align}
with $\lambda_{\text{vae}}$ weighting the VAE loss. Since the GAT architecture is shared, decentralized training across $K$ sites can proceed via local minimization of Eq. \ref{eq:gatloss}, followed by federated averaging of model weights in a server-client setup \cite{FedGAN}.

\textit{\textbf{Tier 2 - Training of reconstruction models:}}
Following Tier 1, each site trains a site-specific reconstruction model $H^k_{\phi^k}$ using its preferred architecture, assisted by the federated GAT prior (Alg.~\ref{alg:site_specific_recon}). To balance within-site and cross-site performance, training proceeds in two phases: pre-training on local data, follow-up training on a hybrid dataset comprising both local and synthetic data. Pre-training uses a standard mean-squared error (MSE) loss \cite{MoDl}:
\begin{equation}
\mathcal{L}^{k}_{\text{rec}}(\mathcal{D}_{\text{loc}}^k) = \mathbb{E}_{(\mathbf{x}^{k}, \mathbf{y}^{k}) \sim \mathcal{D}_{\text{loc}}^k} \left[ \| \mathbf{x}^{k} - H^k_{\phi^k}(\mathcal{F}^{-1}\mathbf{y}^{k},\mathcal{A}^{k}) \|_2^2 \right],
\end{equation}
where $\mathbf{x}^{k}$ is the reference image and $\mathbf{y}^{k}$ the undersampled acquisition from $\mathcal{D}_{\text{loc}}^k$.

For follow-up training, each site generates synthetic coil-combined MR images from all remaining sites using the GAT prior it has received at the end of Tier 1. To simulate multi-coil data, these synthetic images are paired with actual imaging operators sampled from $\mathcal{D}^k_{\text{loc}}$ \cite{Zhu2018,Dar2017}. Specifically, for each sample in $\mathcal{D}_{\text{loc}}^k$, the imaging operator $\mathcal{A}^k = \mathcal{M}^k \mathcal{F} \mathcal{C}^k$ is formed by combining coil sensitivity maps from fully-sampled data ($\mathcal{C}^k$), a random k-space sampling mask ($\mathcal{M}^k$), and Fourier transform ($\mathcal{F}$). This procedure yields synthetic MR images and corresponding multi-coil k-space data at site $k$:
\begin{align}
  &\mathcal{D}^j_{\text{syn}} = \{ \mathbf{x}^j_{\text{syn}},\mathbf{y}^j_{\text{syn}}\} \quad \forall j \in \{1,..,K\} \setminus \{k\}; \\
  &\mbox{ such that: } \mathbf{x}^j_{\text{syn}} = \mathrm{GAT}(\mathbf{sp}(j)),\\
  &\qquad \qquad \,\,\,\, \mathbf{y}^j_{\text{syn}} = \mathcal{A}^k \mathbf{x}^j_{\text{syn}}.
\end{align}
The local MRI dataset and synthetic MRI datasets from remaining sites are then mixed in equal proportions of samples per site in order to construct a hybrid training set:
\begin{equation}
\mathcal{D}^k_{\text{\text{hyb}}} = [\mathcal{D}^k_{\text{\text{loc}}};\, \mathcal{D}_{\text{syn}}^{\{1,..,K\} \setminus \{k\}}].
\end{equation}
The pre-trained model is further trained via the MSE loss:
\begin{equation}
\mathcal{L}^{k}_{\text{rec}}(\mathcal{D}^k_{\text{\text{hyb}}}) = \mathbb{E}_{(\mathbf{x}, \mathbf{y}) \sim \mathcal{D}^k_{\text{hyb}}} \left[ \| \mathbf{x} - H^k_{\phi^k}(\mathcal{F}^{-1}\mathbf{y},\mathcal{A}^{k}) \|_2^2 \right],
\end{equation}
where $\mathbf{x}$ is a reference image and $\mathbf{y}$ is the respective undersampled acquisition drawn from $\mathcal{D}_{\text{hyb}}^k$.

\section{Methods}
\subsection{Implementation Details for FedGAT}
GAT was adapted to multi-site MRI data from a foundation model designed for natural images \cite{VAR}. $S\,$=$\,10$ spatial scales were used, each of size $h_s$=$w_s$=$p$ where $p \in {1, 2, 3, 4, 5, 6, 8, 10, 13, 16}$. This configuration balanced spatial granularity and computational load, enabling progressive modeling of anatomical structure from coarse to fine detail. Both the VAE encoder and decoder comprised 6 stages with 12 residual convolutional blocks; the encoder used stride-2 convolutions for downsampling, and the decoder employed transposed convolutions for upsampling. A channel dimensionality of $c\,$=$\,32$ and a codebook size of $V\,$=$\,4096$ were used. The transformer module contained $L\,$=$\,16$ sequential blocks, each with embedding dimensionality $d\,$=$\,1024$ and 16 self-attention heads. VAE modules were frozen after initialization from the pre-trained `vae\textunderscore ch160v4096z32' model\footnote{\url{https://huggingface.co/FoundationVision/var/resolve/main/vae_ch160v4096z32.pth}}. Transformer weights were adapted together with the randomly initialized prompts and gMLP for the site-prompting mechanism.

\subsection{Competing Methods}
FedGAT was evaluated against non-FL benchmarks (Central \cite{kaissis2020secure}, Single \cite{guo2021}), as well as distillation-based (FedDF \cite{feddf}, FedMD \cite{fedmd}) and generative (FedGIMP \cite{elmas2022federated}, FedDDA \cite{FedDDA}) FL baselines. \textit{As prior FL methods for MRI reconstruction were not tailored for model-heterogeneous settings,} we either adopted model-agnostic baselines from non-reconstruction domains (FedMD, FedDF, FedDDA) or adapted reconstruction baselines to support architectural heterogeneity (FedGIMP adopted to follow the two-tier strategy in FedGAT). As model-heterogeneous settings preclude aggregation via simple weight averaging, FL baselines used either model distillation or synthetic image generation as a means for cross-site knowledge transfer. Note that distillation-based aggregation involves bidirectional distillation between global and site-specific models, each direction incurring considerable computational overhead, as the student models must be trained using a substantial set of teacher-generated input-output sample pairs. Here, we observed that distillation-based baselines (FedDF, FedMD) attained a near-optimal trade-off between performance and efficiency with a total of 200 training epochs and distillation every 100 epochs. In generative baselines (FedGIMP, FedDDA), the two-tier strategy in FedGAT was followed under matching training procedures with $N_c$$\,=\,$500, $N_l$$\,=\,$1, $N_p$$\,=\,$100 and $N_f$$\,=\,$100.

\subsection{Modeling Procedures}
We considered multiple three-site FL setups to evaluate performance under model-heterogeneous settings, using reconstruction models including MoDL \cite{MoDl}, rGAN \cite{rgan}, and D5C5 \cite{Schlemper2017}. All sites used a common acceleration rate and sampling density. In the first experiment, each site employed a unique architecture—either adversarial or unrolled (MoDL, rGAN, D5C5)—while in the second, all the sites trained the same type of physics-driven model, albeit using different numbers of cascades in the unrolled backbone  (MoDL3, MoDL5, MoDL7). An unrolled TransUNet architecture \cite{chen2021transunet} was adopted for experiments involving a non-participating site. All models were implemented in PyTorch and trained on RTX 4090 GPUs using the AdamW optimizer with $\beta_1 = 0.9$, $\beta_2 = 0.95$, weight decay of 0.05, and learning rate $\eta = 10^{-4}$. Reconstruction performance was assessed using peak signal-to-noise ratio (PSNR) and structural similarity (SSIM), while the fidelity of synthetic MR images was measured using the Fréchet Inception Distance (FID). Statistical significance of performance differences among competing methods was determined via Wilcoxon signed-rank tests (p$<$0.05).

\input{tables/table8.tex}

\begin{figure}[t] 
\centering
\includegraphics[width=0.975\columnwidth]{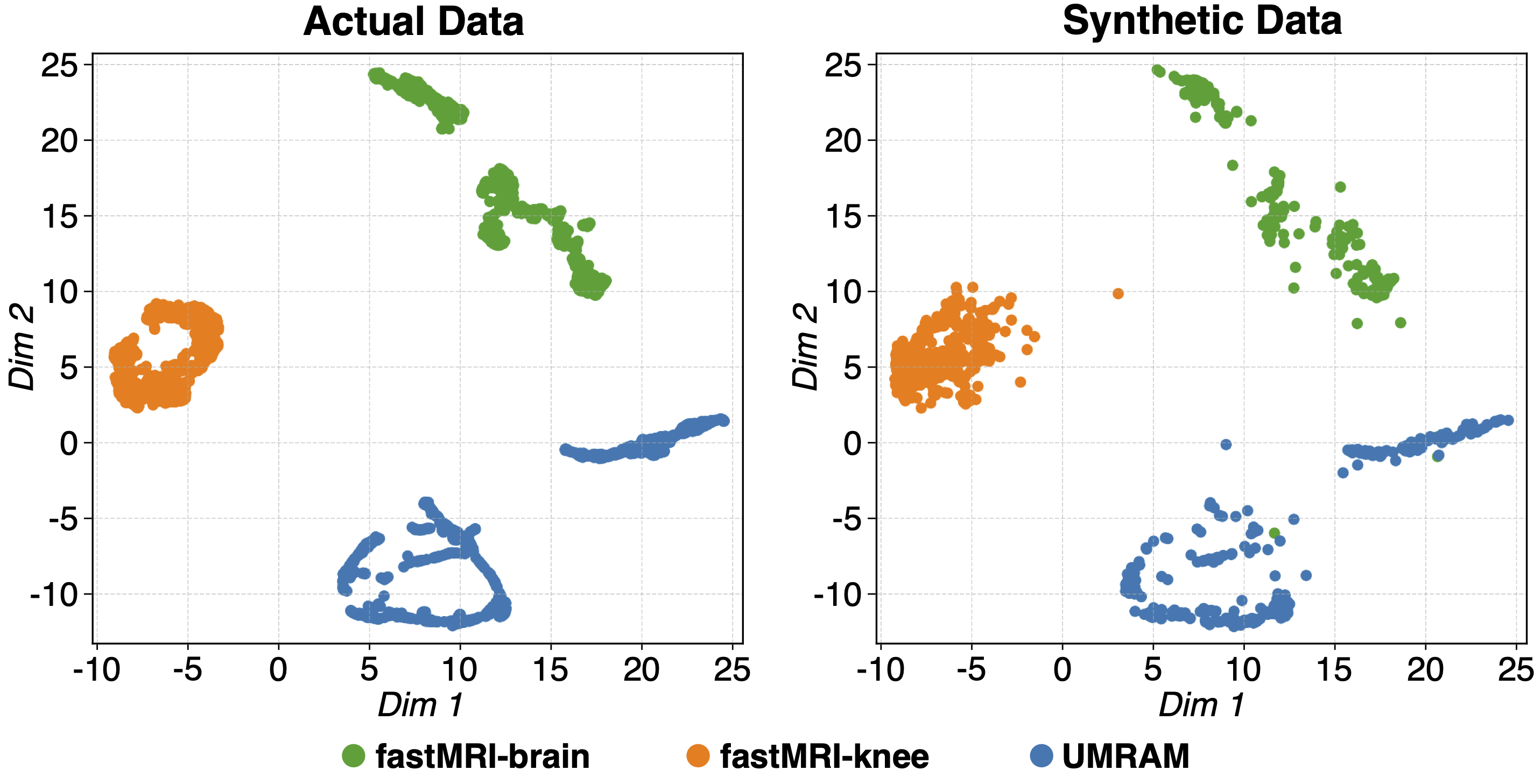}
\caption{
UMAP projection of deep feature embeddings from actual (left) and synthetic (GAT prior; right) MR images across three sites (fastMRI-knee, fastMRI-brain, UMRAM).
}
\label{fig:umap}
\vspace{-0.25cm}
\end{figure}

\input{tables/table1.tex}
\input{tables/table2.tex}

\subsection{Datasets}
Multi-contrast k-space data from public fastMRI-brain, fastMRI-knee \cite{fastMRI} and an in-house dataset were analyzed \cite{elmas2022federated}. Experimental procedures for collecting the in-house dataset were approved by the local ethics committee at Bilkent University, and written informed consent was obtained from all participants. In experiments involving a non-participating site, multi-coil k-space data from the Calgary dataset \cite{calgary} were analyzed. Each dataset was assigned to a separate site, and across sites, (training, validation, test) splits of (108, 18, 24) non-overlapping subjects were used, resulting in (2520, 315, 630) cross-sectional images. Multi-coil data were compressed to 5 virtual coils \cite{Zhang2013}. Both actual and synthetic k-space data were retrospectively undersampled using a variable-density (VD) pattern with acceleration factors R = 4x-8x \cite{Lustig2007}. Models were trained on a mixture of multi-contrast MRI datasets.

\section{Results}

\input{tables/table3.tex}
\input{tables/table4.tex}

\subsection{Fidelity of the Federated GAT Prior}
The GAT prior is central to our framework as it generates the synthetic multi-site MR images used for training reconstruction models. Since the quality of training data directly influences model effectiveness, we first evaluated the fidelity of GAT. For qualitative assessment, we visualized deep feature embeddings from actual and synthetic MR images at each site, using a UMAP manifold (Fig. \ref{fig:umap}). We find that synthetic samples closely align with actual ones, suggesting that the prior preserves both within-site and across-site structure.

We further quantified the fidelity of synthetic images generated via the GAT prior (Table \ref{tab:fid-crosssite}). FID scores were computed both between pairs of actual samples (to capture reference relationships among sites) and between synthetic–actual pairs (to evaluate generated data). Afterwards, we derived two Spearman correlation–based measures: a within-site consistency measure, obtained by correlating the synthetic–actual FID matrix with an idealized identity matrix (zeros on the diagonal, ones off-diagonal), and an across-site structure preservation measure, obtained by correlating the off-diagonal entries of synthetic–actual and actual–actual FID matrices. High correlation values (within-site consistency = 0.92±0.04; cross-site preservation = 0.98±0.04) confirm that the GAT prior produces synthetic data that faithfully reflect both within-site and across-site structure of multi-site MRI data distributions.

\subsection{Comparison Studies}
We evaluated FedGAT in model-heterogeneous FL settings against distillation (FedDF, FedMD) and generative (FedGIMP, FedDDA) baselines. Within-site performance was measured by deploying a site-specific model on its local data. Across-site performance was measured by deploying the site-specific model on data from other sites, and averaging performance across those external sites. For benchmarking, `Single' models trained exclusively on single-site data, and `Central' models trained on data pooled across sites (a privacy-violating benchmark) were included. Two FL scenarios were considered: one with entirely distinct model architectures across sites (MoDL, rGAN, D5C5; Table~\ref{tab:hetA_multi}) and another with variants of a MoDL architecture of varying complexity (3, 5, or 7 cascades; Table~\ref{tab:hetB_multi}).

\ul{\textit{Within-site reconstructions:}} FedGAT significantly outperforms all FL baselines in each scenario (p$<$0.05). Averaged across both scenarios, it improves (PSNR, SSIM) by (1.4dB, 0.6\%) over distillation and (2.1dB, 1.9\%) over generative baselines. It also surpasses `Single' by (2.2dB, 1.2\%) and is the closest FL method to `Central'. These findings suggest that FedGAT effectively leverages multi-site MRI data generated via the GAT prior to build performant site-specific reconstruction models, yielding strong in-domain image quality without compromising local specialization.

\begin{figure*}[t]
\centering
\begin{minipage}[t]{0.25\textwidth}
\captionsetup{justification=justified, width=\linewidth}
\vspace{45pt}
\caption{Representative reconstructions at R=8x from zero-filled Fourier method (Zero-filled), single-site models (Single), FL baselines (FedDF, FedMD, FedGIMP, FedDDA), and FedGAT, along with reference images.
\textbf{(a)} Site-specific fastMRI-knee model tested on UMRAM, 
\textbf{(b)} Site-specific fastMRI-brain model tested on fastMRI-knee, 
\textbf{(c)} Site-specific UMRAM model tested on fastMRI-brain. 
Zoom-in windows of error maps and images are included to emphasize method differences.}
\label{fig:visualresB}
\end{minipage}\hfill
\begin{minipage}[t]{0.67\textwidth}
\vspace{0pt}
\includegraphics[width=\linewidth]{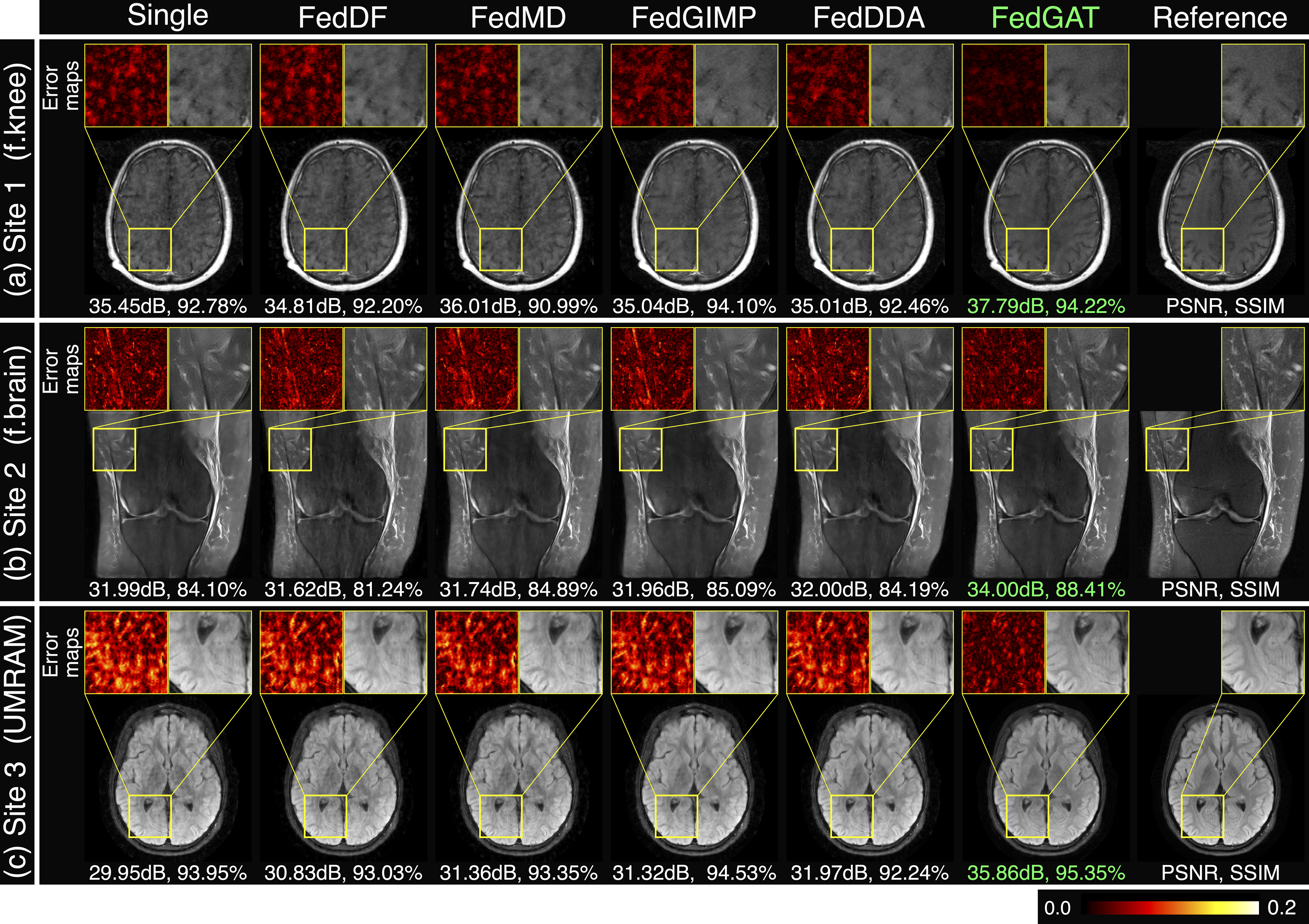}
\end{minipage}
\end{figure*}

\ul{\textit{Across-site reconstructions:}} FedGAT again yields significant gains in each scenario (p$<$0.05), offering (1.7dB, 0.9\%) improvement over distillation and (3.3dB, 3.0\%) over generative baselines. Meanwhile, it exceeds `Single' by (2.9dB, 1.9\%), while closely approaching `Central''s performance. Representative images reconstructed are depicted in Fig.~\ref{fig:visualresB} for the second scenario. FedGAT yields higher image quality compared to baselines, with sharper depiction of structural details and reduced artifacts and noise. These results suggest that FedGAT enables effective knowledge transfer across participating sites by capturing shared anatomical patterns and site-specific data characteristics, supporting robust reconstruction despite differences in models and datasets.

\subsection{Extension to a Non-Participating Site}
\begin{figure}[t]
\centering
\includegraphics[width=0.85\columnwidth]{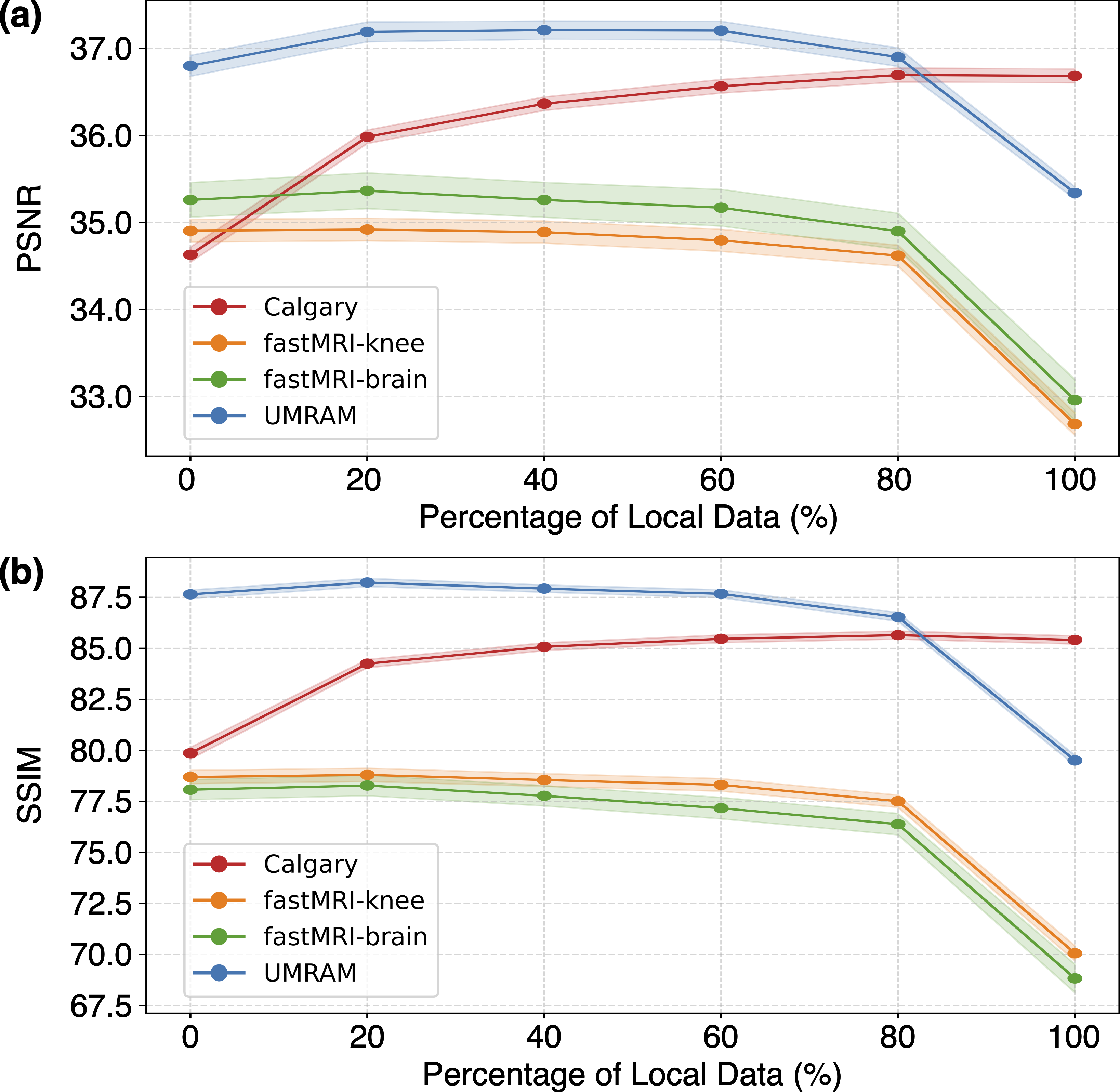}
\caption{Performance of site-specific models trained at a held-out site (Calgary) on hybrid datasets combining local and GAT-generated synthetic data from a three-site FL setup (fastMRI-knee, fastMRI-brain, UMRAM). With the proportion of local data systematically varied, results reflect within-site (Calgary) and and across-site reconstruction performance at the FL sites (see legend).}
\label{fig:nonparticipating}
\vspace{-0.25cm}
\end{figure}

FedGAT enables site-specific models to generalize effectively across participating sites under model heterogeneity. Importantly, it can also boost generalization performance of models built at non-participating sites through the federated GAT prior. To assess this, the GAT prior trained using a three-site FL setup (fastMRI-knee, fastMRI-brain, UMRAM) was deployed to train an unrolled transformer model at a held-out site (Calgary). At the held-out site, reconstruction models were trained using varying proportions of local and synthetic data ranging from 0\% local / 100\% synthetic (i.e., P0 model) to 100\% local / 0\% synthetic (i.e., P100 model). Trained models were used to reconstruct images at the held-out site and also at each individual participating site, as shown in Fig. \ref{fig:nonparticipating}.

\ul{\textit{Within-site reconstructions at the held-out site:}} Reconstruction performance at Calgary shows that while the P0 model moderately underperforms the P100 model (–2.0dB PSNR, –5.5\% SSIM), mixed-data models (trained on 40–80\% local data) close this gap ($\sim$ -0.1dB PSNR, -0.02\% SSIM difference). This indicates that synthetic data from FedGAT enables the development of highly performant reconstruction models, even when local training data is limited.

\ul{\textit{Across-site reconstructions at participating sites:}} Reconstruction performances at fastMRI-knee, fastMRI-brain and UMRAM reveal that the P100 model (100\% local data) generalizes poorly, while mixed-data models (trained on 20–80\% local data) achieve better generalization—gaining 2.1dB PSNR and 8.3\% SSIM on average over the P100 model. This demonstrates that FedGAT-generated synthetic data can enhance the cross-site robustness of reconstruction models developed at a non-participating institution.

\input{tables/table5.tex}
\input{tables/table6.tex}

\input{tables/table7.tex}

\subsection{Ablation Studies}
\subsubsection{Privacy}
As an FL technique, FedGAT communicates the parameters of the GAT prior rather than raw MRI data, thereby mitigating patient privacy risks. Although generative priors may still be susceptible to cross-site information leakage \cite{kaissis2020secure}, we reasoned that incorporating a site-specific prompting mechanism could help alleviate these risks. To evaluate this, we followed the framework in \cite{andrew2023one}, introducing an intruder site with white Gaussian noise into the original three-site FL setup (fastMRI-knee, fastMRI-brain, UMRAM). We then evaluated the intruder's impact on the fidelity of synthetic images at the original sites by comparing GAT priors trained with and without site prompts. Table~\ref{tab:fid-delta} shows that the non-prompted variant suffers a noticeable fidelity drop, whereas the site-prompted version remains more robust to the intrusion, underscoring its role in enhancing resilience against information leakage.

\subsubsection{Key Design Elements}
Next, we conducted ablation studies to evaluate the impact of key design elements in FedGAT. To assess the role of the learnable site prompt implemented via a gated MLP, we compared a `w one-hot prompt' variant using a fixed one-hot site index and a `w/o site prompt' variant that removed the prompt entirely from the transformer input. To examine the contribution of multi-scale autoregressive prediction, a `w/o scales' variant set $S=1$ to disable multi-scale generation, while a `w/o autoreg' variant omitted $f_{<S}$, preventing cross-scale dependency during token map prediction. To evaluate the utility of hybrid training on augmented datasets for site-specific reconstruction models, we implemented a `w model agg' variant using conventional model aggregation \cite{FedAvg}, and a `w synthetic' variant that trained reconstruction models solely on synthetic data. All comparisons were conducted under model-homogeneous settings to eliminate confounding effects from architectural variability. As summarized in Table~\ref{tab:ablations}, FedGAT consistently outperforms all variants (p$<$0.05), underscoring the importance of its learned site prompt, multi-scale autoregressive token generation, and hybrid local-synthetic training in achieving superior reconstruction performance.

\section{Discussion}
We introduced a model-agnostic FL framework for MRI reconstruction that supports heterogeneous architectures by decoupling cross-site transfer of knowledge regarding MRI data distributions from local training of reconstruction models. FedGAT achieves this by leveraging a federated GAT prior that generates MR images via autoregressive predictions across multiple spatial scales, guided by a site prompt for controllable synthesis. Prior efforts to construct generative priors have relied on adversarial models prone to training instability and poor image quality \cite{rgan,Mardani2019b}, or on diffusion models with long inference times and residual noise \cite{jalaln2021nips}. In contrast, the autoregressive design of GAT offers improved fidelity in FL settings, as suggested by our results. To locally build an augmented multi-site dataset, the federated GAT prior is deployed at each site to synthesize MR images from remaining sites, enabling site-specific reconstruction models of different architectures across sites to be trained independently. This approach promotes generalization while preserving architectural freedom, offering a scalable and practical pathway for FL in multi-institutional MRI studies.

FedGAT offers a computationally efficient alternative to conventional FL by decoupling generative modeling from reconstruction. Traditional FL approaches in MRI require consensus on a shared architecture for the reconstruction model, and incur substantial overhead when retraining is triggered by architectural changes \cite{guo2021}. In contrast, FedGAT trains a federated prior once on multi-site MRI data and can reuse it to synthesize training data for building local reconstruction models of varying architectures, thereby amortizing training costs across architectural experiments. Unlike distillation-based FL methods that require frequent alignment between global and local networks \cite{feddf}, FedGAT facilitates knowledge transfer through synthetic images generated after the federated learning phase, thereby alleviating communication overhead. Compared to prior generative FL methods using adversarial or diffusion priors \cite{FedDDA}, FedGAT strikes a better balance between sample fidelity and inference cost—achieving higher image quality than both, while being faster than diffusion models (10 autoregressive stages vs. 1000 steps) and moderately slower than adversarial approaches that require just a single step.

While FL frameworks improve patient privacy by sharing model weights rather than raw MRI data, they remain susceptible to adversarial attacks that corrupt updates and degrade reconstruction fidelity \cite{kaissis2020secure}, as well as inference attacks that extract sensitive information from trained weights. Conventional FL methods that transmit global reconstruction model parameters are particularly exposed to such risks. In contrast, FedGAT enhances security by constructing reconstruction models locally without sharing their weights, offering greater resilience to both training- and inference-time threats. Potential risks associated with the global GAT prior may be mitigated through differential privacy mechanisms during training \cite{FengICCV2021}, though further work is needed to systematically assess FedGAT’s privacy-preserving properties.

\section{Conclusion}
Here, we introduced a novel model-agnostic federated learning technique that enables multiple sites with distinct architectural preferences to collaborate in building MRI reconstruction models. To support model-heterogeneous settings, FedGAT decouples a first tier to learn a federated generative prior that captures the distribution of multi-site MR images, from a second tier to build site-specific albeit generalizable reconstruction models with the aid of synthetic datasets locally generated by the federated prior. Experimental results show that FedGAT outperforms state-of-the-art FL baselines in both within-site and across-site reconstructions, highlighting its ability to transfer knowledge effectively while remaining architecture-agnostic—thus addressing a key challenge in enabling flexible, multi-institutional collaborations.

\bibliographystyle{IEEETran}
\bibliography{Papers}

\end{document}

%% file: tables/table8.tex
\begin{table}[t]
\centering
\setlength\tabcolsep{6pt}
\renewcommand{\arraystretch}{1.4}
\caption{
FID scores between source and target sites. A$\to$A: actual source-site images vs. actual target-target images. S$\to$A: synthetic source-site images vs. actual target-site images.
}
\resizebox{0.95\columnwidth}{!}{%
\begin{tabular}{|l|l|c|c|c|c|c|c|c|}
\cline{3-8}
\multicolumn{2}{c|}{\textbf{}} & \multicolumn{6}{c|}{{\textit{Target site}}} \\
\cline{3-8}
\multicolumn{2}{c|}{\textbf{}} & \multicolumn{2}{c|}{{fastMRI-knee}} & \multicolumn{2}{c|}{{fastMRI-brain}} & \multicolumn{2}{c|}{{UMRAM}} \\
\cline{3-8}
\multicolumn{2}{c|}{\textbf{}}  & A$\rightarrow$A & S$\to$A & A$\to$A & S$\to$A & A$\to$A & S$\to$A \\
\hline
\multirow{3}{*}{\rotatebox[origin=c]{90}{\textit{Source site}}} & fastMRI-knee   & 0.00 & 0.65 & 2.04 & 1.78 & 10.18 & 9.39\\
& fastMRI-brain  &  2.04 & 2.63 & 0.00 & 0.54 & 3.81 & 3.91 \\
& UMRAM          & 10.18 & 9.66 & 3.81 & 3.51 & 0.00 & 0.07\\
\hline
\end{tabular}%
}
\label{tab:fid-crosssite}
\end{table}

%% file: tables/table2.tex
\begin{table}[t]
\centering
\caption{Within-site and across-site reconstruction performance (Site 1: fastMRI-knee, Site 2: fastMRI-brain, Site 3: UMRAM). Each site employs a distinct model type. \underline{Underlining} indicates non-FL benchmarks. Boldface indicates the top-performing FL method at each site and acceleration factor.}
\resizebox{1.0\columnwidth}{!}{%
\begin{tabular}{| l l | c | c | c | c | c | c | c | c |}
 \hline
\multicolumn{2}{|c|}{\textbf{Within-site}} & \multicolumn{2}{c|}{Site 1 (MoDL)} & \multicolumn{2}{c|}{Site 2 (rGAN)} & \multicolumn{2}{c|}{Site 3 (D5C5)} &  \multicolumn{2}{c|}{Average}\\
\cline{3-10}
\multicolumn{2}{|c|}{\textbf{recon}} & PSNR & SSIM & PSNR & SSIM & PSNR & SSIM & PSNR & SSIM  \\
 \hline
\multirow{2}{*}{\underline{Single}}& 4x  &   40.0$\pm$3.1 &96.2$\pm$1.5& 38.3$\pm$3.3&97.8$\pm$1.1&  34.8$\pm$1.9 & 94.7$\pm$1.5 & 37.3 & 96.2\\
& 8x    &  29.4$\pm$4.1 &  84.5$\pm$5.1& 35.0$\pm$3.1 & 95.4$\pm$1.9 &31.2$\pm$2.1 & 90.6$\pm$2.6&  31.9 & 90.2  \\
\hline
\multirow{2}{*}{\underline{Central}} & 4x   &  40.0$\pm$3.0 &  96.2$\pm$1.5 & 40.4$\pm$3.0 & 98.0$\pm$0.9 & 37.0$\pm$1.7 & 95.8$\pm$1.3 &  39.1 & 96.7  \\
& 8x   &  35.8$\pm$2.7 & 90.2$\pm$4.2 & 36.2$\pm$3.1 & 95.8$\pm$1.7 & 34.1$\pm$1.7 & 93.6$\pm$2.1 & 35.4 & 93.2 \\
\hline
\multirow{2}{*}{FedDF}  & 4x   &  39.2$\pm$3.2&  96.1$\pm$1.5 & 38.6$\pm$3.5 & 97.5$\pm$1.4& 36.0$\pm$1.6 & 94.2$\pm$2.0& 37.9 & 95.9\\
& 8x    &   34.6$\pm$3.3 &89.7$\pm$4.3 &35.2$\pm$3.4 & 95.0$\pm$2.4& 31.8$\pm$2.2&90.6$\pm$3.8& 33.9 & 91.8\\
\cline{1-10}
\multirow{2}{*}{FedMD} & 4x   & 38.4$\pm$3.3 &  96.0$\pm$1.4&  38.2$\pm$3.4 &  97.8$\pm$1.0 & 35.5$\pm$1.6 & 94.0$\pm$1.5& 37.4 & 95.9 \\
& 8x    &   34.7$\pm$3.2 &  89.4$\pm$4.5 &34.8$\pm$3.4&95.3$\pm$2.0 & 31.5$\pm$2.3 &90.8$\pm$2.3& 33.7 & 91.8\\ 
\cline{1-10}
\multirow{2}{*}{FedGIMP}  & 4x &39.6$\pm$3.3 & 96.3$\pm$1.5 & 39.4$\pm$3.4 & 97.9$\pm$0.9 & 36.0$\pm$1.6& 94.8$\pm$1.5  & 38.3 &  96.3\\
 & 8x & 34.9$\pm$3.1 &90.1$\pm$4.3& 35.4$\pm$3.3 & 95.2$\pm$2.0 &31.8$\pm$2.0 & 89.4$\pm$3.0 & 34.0 & 91.6\\
\cline{1-10}
\multirow{2}{*}{FedDDA} & 4x & 37.5$\pm$3.6 &95.0$\pm$1.5& 38.3$\pm$3.3 & 97.7$\pm$1.0 & 31.1$\pm$2.5 & 86.1$\pm$4.1 & 35.6 & 92.9 \\
 & 8x  &  34.1$\pm$3.3 & 89.4$\pm$3.5& 33.3$\pm$3.4 & 94.0$\pm$2.3 & 27.9$\pm$2.6 & 80.9$\pm$5.1 & 31.8 & 88.1 \\
\cline{1-10}
\multirow{2}{*}{FedGAT}  & 4x   & \textbf{40.1$\pm$3.2} & \textbf{96.4$\pm$1.5}   & \textbf{40.2$\pm$3.2} & \textbf{98.0$\pm$0.9} &  \textbf{36.1$\pm$1.5} & \textbf{94.9$\pm$1.4} & \textbf{38.8} &  \textbf{96.4}\\
 & 8x &   \textbf{35.5$\pm$3.0} & \textbf{90.5$\pm$4.2}&  \textbf{35.6$\pm$3.1} & \textbf{95.4$\pm$1.9} & \textbf{32.1$\pm$1.9}& \textbf{90.9$\pm$2.4} & \textbf{34.4} & \textbf{92.3} \\
 \hline
 \multicolumn{10}{c}{\mbox{ }} \\[-1.75ex]
 \hline
\multicolumn{2}{|c|}{\textbf{Across-site}} & \multicolumn{2}{c|}{Site 1 (MoDL)} & \multicolumn{2}{c|}{Site 2 (rGAN)} & \multicolumn{2}{c|}{Site 3 (D5C5)} &  \multicolumn{2}{c|}{Average}\\
\cline{3-10}
\multicolumn{2}{|c|}{\textbf{recon}} & PSNR & SSIM & PSNR & SSIM & PSNR & SSIM & PSNR & SSIM  \\
 \hline
 \multirow{2}{*}{\underline{Single}}& 4x &   44.9$\pm$3.9 & 98.9$\pm$0.7 & 33.4$\pm$3.7 & 94.9$\pm$3.4 &33.1$\pm$2.5& 89.4$\pm$5.2& 37.1 & 94.4 \\
 & 8x &  28.2$\pm$4.0 & 89.5$\pm$4.2 & 29.6$\pm$4.4 & 88.7$\pm$9.0 &29.5$\pm$2.6 & 81.9$\pm$8.7& 29.1  & 86.7 \\
\cline{1-10}
\multirow{2}{*}{\underline{Central}} & 4x &  46.6$\pm$3.8& 99.0$\pm$0.6 & 40.2$\pm$3.6& 97.0$\pm$2.4 & 33.8$\pm$2.9 & 88.9$\pm$6.2& 40.2 & 95.0 \\
 & 8x &  40.6$\pm$3.0 & 97.4$\pm$1.4 & 35.9$\pm$3.6 & 92.9$\pm$5.8 & 30.6$\pm$2.8 &82.4$\pm$8.9& 35.7 & 90.9 \\ 
\hline 
\multirow{2}{*}{FedDF} & 4x &  44.7$\pm$4.7 & 98.9$\pm$0.7 & 31.7$\pm$4.5 & 93.6$\pm$4.9& 33.2$\pm$2.8& 88.0$\pm$6.7& 36.5 & 93.5\\
 & 8x &  38.3$\pm$4.2 & 96.9$\pm$1.8 &  29.3$\pm$4.0& 88.8$\pm$8.3&  29.4$\pm$2.8&  80.8$\pm$10.4& 32.3& 88.8 \\
\hline
\multirow{2}{*}{FedMD} & 4x &  45.0$\pm$4.5 & 98.9$\pm$0.7 & 33.8$\pm$4.4&  94.9$\pm$3.8 &  33.0$\pm$2.7& 89.1$\pm$4.7 & 37.3 & 94.3 \\
 & 8x &  38.7$\pm$4.1 & 96.8$\pm$1.8 & 30.9$\pm$4.2 & 89.4$\pm$8.7 & 29.7$\pm$2.7 & 82.4$\pm$7.9 & 33.1 & 89.5 \\
\hline
\multirow{2}{*}{FedGIMP} & 4x & 42.8$\pm$4.4 &  98.9$\pm$0.7& 31.1$\pm$3.8& 93.0$\pm$3.1 &33.4$\pm$2.5 & 89.8$\pm$5.3& 35.8 & 93.9  \\
 & 8x &  37.7$\pm$3.7& 97.2$\pm$1.5& 28.7$\pm$2.9 &87.1$\pm$7.2 & 29.3$\pm$2.6 & 82.3$\pm$8.4 &31.9  &  88.9\\
\hline
\multirow{2}{*}{FedDDA} & 4x & 37.5$\pm$5.2 & 95.3$\pm$2.5& 32.9$\pm$2.8 & 92.7$\pm$2.8 &28.5$\pm$2.8 & 78.2$\pm$8.4 & 33.0 & 88.7 \\
 & 8x &  34.4$\pm$4.3 & 93.2$\pm$2.9 & 29.3$\pm$2.9 & 85.2$\pm$6.8& 26.6$\pm$2.6 & 70.9$\pm$11.1& 30.1 &  83.1\\
\hline 
\multirow{2}{*}{FedGAT} & 4x & \textbf{45.5$\pm$4.2} & \textbf{99.1$\pm$0.7}  & \textbf{37.0$\pm$4.5} & \textbf{95.4$\pm$3.3}& \textbf{33.9$\pm$2.6}&\textbf{89.9$\pm$5.1} & \textbf{38.8} & \textbf{94.8} \\
 & 8x &\textbf{38.8$\pm$3.6} & \textbf{97.5$\pm$1.5} &  \textbf{32.6$\pm$4.0} & \textbf{90.8$\pm$6.7} & \textbf{30.3$\pm$2.7} &\textbf{83.4$\pm$8.0}& \textbf{33.9} & \textbf{90.6}  \\
\hline
\end{tabular}%
}
\label{tab:hetA_multi}
\vspace{-0.25cm}
\end{table}

%% file: tables/table4.tex
\begin{table}[t]
\centering
\caption{Within-site and across-site reconstruction performance (Site 1: fastMRI-knee, Site 2: fastMRI-brain, Site 3: UMRAM). Each site employs an MoDL variant with different number of cascades.}
\label{tab:hetB_multi}
\resizebox{\columnwidth}{!}{%
\begin{tabular}{| l l | c | c | c | c | c | c | c | c |}
 \hline
\multicolumn{2}{|c|}{\textbf{Within-site}} & \multicolumn{2}{c|}{Site 1 (MoDL3)} & \multicolumn{2}{c|}{Site 2 (MoDL5)} & \multicolumn{2}{c|}{Site 3 (MoDL7)} & \multicolumn{2}{c|}{Average} \\
\cline{3-10}
\multicolumn{2}{|c|}{\textbf{recon}} & PSNR & SSIM & PSNR & SSIM & PSNR & SSIM & PSNR & SSIM \\
 \hline
 \multirow{2}{*}{\underline{Single}}& 4x  
& 40.0$\pm$3.1 & 96.2$\pm$1.5
& 45.8$\pm$3.7 & 98.8$\pm$0.5
& 41.4$\pm$4.2 & 99.0$\pm$0.6
& 42.4 & 98.0
\\
& 8x   
& 29.4$\pm$4.1 & 84.5$\pm$5.1
& 39.7$\pm$3.2 & 96.5$\pm$1.2
& 41.1$\pm$3.2 & 98.4$\pm$0.9
& 36.7 & 93.1
\\
\hline
\multirow{2}{*}{\underline{Central}} & 4x   
& 40.0$\pm$3.0 & 96.2$\pm$1.5
& 46.9$\pm$4.0 & 99.1$\pm$0.6
& 52.2$\pm$3.1 & 99.8$\pm$0.1
& 46.4 & 98.4
\\
& 8x   
& 35.8$\pm$2.7 & 90.2$\pm$4.2
& 40.5$\pm$3.1 & 97.0$\pm$1.2
& 45.1$\pm$2.3 & 99.0$\pm$0.5
& 40.5 & 95.4
\\
\hline
\multirow{2}{*}{FedDF}  & 4x 
& 38.8$\pm$3.5 & 95.9$\pm$1.7 & 43.5$\pm$4.8 & 98.5$\pm$0.9 & 45.8$\pm$5.6 & 99.5$\pm$0.4 & 42.7 & 98.0\\
& 8x    &  34.1$\pm$3.5 & 89.0$\pm$4.7 & 37.4$\pm$3.7 & 95.4$\pm$1.6 & 39.0$\pm$4.7 & 98.1$\pm$1.1 & 36.8 & 94.2\\
\hline
\multirow{2}{*}{FedMD} & 4x & 38.4$\pm$3.4 & 96.0$\pm$1.4 & 44.5$\pm$4.2 & 98.6$\pm$0.6 & 45.5$\pm$6.1 & 99.5$\pm$0.4 & 42.8 & 98.0\\
 & 8x & 34.7$\pm$3.2 & 89.5$\pm$4.5 & 38.9$\pm$3.7 & 95.9$\pm$1.6 & 38.9$\pm$5.1 & 98.2$\pm$1.2 & 37.5 & 94.5\\
\hline
\multirow{2}{*}{FedGIMP}  & 4x & 39.6$\pm$3.3 & 96.3$\pm$1.5 & 45.3$\pm$3.4 & 98.6$\pm$0.6 & 45.0$\pm$3.4 & 99.3$\pm$0.4 & 43.3 & 98.1\\
& 8x & 34.9$\pm$3.1 & 90.1$\pm$4.3 & 39.5$\pm$3.2 & 96.4$\pm$1.5 & 40.5$\pm$2.5 & 98.4$\pm$0.7 & 38.3 & 95.0\\
\hline
\multirow{2}{*}{FedDDA} & 4x & 37.5$\pm$3.6 & 95.0$\pm$1.5 & 42.7$\pm$3.3 & 97.4$\pm$1.3 & 38.6$\pm$3.6 & 93.2$\pm$1.8 & 39.6 & 95.2\\
& 8x & 34.1$\pm$3.3 & 89.4$\pm$3.5 & 38.6$\pm$3.1 & 95.6$\pm$1.7 & 36.1$\pm$3.6 & 91.8$\pm$2.1 & 36.3 & 92.3\\
\hline
\multirow{2}{*}{FedGAT}  & 4x & \textbf{40.1$\pm$3.2} & \textbf{96.4$\pm$1.5} & \textbf{45.8$\pm$3.6} & \textbf{98.7$\pm$0.6} & \textbf{47.1$\pm$3.7} & \textbf{99.6$\pm$0.3} & \textbf{44.3} & \textbf{98.2}\\
& 8x & \textbf{35.5$\pm$3.0} & \textbf{90.5$\pm$4.2} & \textbf{39.9$\pm$3.3} & \textbf{96.8$\pm$1.4} & \textbf{42.7$\pm$3.0} & \textbf{98.7$\pm$0.7} & \textbf{39.4} & \textbf{95.3}\\
 \hline
 \multicolumn{10}{c}{\mbox{ }} \\[-1.75ex]
 \hline
\multicolumn{2}{|c|}{\textbf{Across-site}} & \multicolumn{2}{c|}{Site 1 (MoDL3)} & \multicolumn{2}{c|}{Site 2 (MoDL5)} & \multicolumn{2}{c|}{Site 3 (MoDL7)} & \multicolumn{2}{c|}{Average} \\
\cline{3-10}
\multicolumn{2}{|c|}{\textbf{recon}} & PSNR & SSIM & PSNR & SSIM & PSNR & SSIM & PSNR & SSIM  \\
 \hline
\multirow{2}{*}{\underline{Single}}& 4x
& 44.9$\pm$3.9 & 98.9$\pm$0.7 & 44.2$\pm$4.3 & 98.2$\pm$1.6 & 40.0$\pm$3.6 & 96.9$\pm$2.0 & 43.0 & 98.0\\
  & 8x 
& 28.2$\pm$4.0 & 89.5$\pm$4.2 & 38.6$\pm$4.0 & 94.6$\pm$4.9 & 37.4$\pm$3.5 & 93.6$\pm$3.8 & 34.7 & 92.5\\
\hline
\multirow{2}{*}{\underline{Central}} & 4x 
& 46.6$\pm$3.8 & 99.0$\pm$0.6 & 46.5$\pm$5.6 & 98.6$\pm$1.5 & 45.0$\pm$4.4 & 98.4$\pm$1.1 & 46.0 & 98.7\\
 & 8x 
& 40.6$\pm$3.0 & 97.4$\pm$1.4 & 40.1$\pm$4.2 & 95.1$\pm$4.7 & 39.1$\pm$3.7 & 94.5$\pm$3.7 & 39.9 & 95.7\\
 \hline 
\multirow{2}{*}{FedDF} & 4x & 42.1$\pm$5.1 & 98.7$\pm$0.8 & 40.2$\pm$5.2 & 97.7$\pm$2.0 & 42.1$\pm$5.0 & 97.6$\pm$1.5 & 41.5 & 98.0\\
  & 8x & 35.6$\pm$4.2 & 96.3$\pm$1.9 & 35.2$\pm$4.3 & 93.6$\pm$5.3 & 36.4$\pm$3.9 & 92.7$\pm$4.2 & 35.7 & 94.2\\
\hline
\multirow{2}{*}{FedMD} & 4x & 45.2$\pm$4.5 & 98.9$\pm$0.7 & 42.1$\pm$5.9 & 98.0$\pm$1.8 & 42.7$\pm$4.8 & 97.9$\pm$1.4 & 43.3 & 98.3\\
  & 8x & 38.7$\pm$4.2 & 96.9$\pm$1.8 & 37.0$\pm$4.6 & 94.1$\pm$5.1 & 36.7$\pm$4.2 & 93.1$\pm$4.2 & 37.5 & 94.7\\
\hline
\multirow{2}{*}{FedGIMP} & 4x & 42.8$\pm$4.4 & 98.9$\pm$0.7 & 42.9$\pm$4.7 & 97.9$\pm$1.8 & 40.9$\pm$4.0 & 97.6$\pm$1.6 & 42.2 & 98.1\\
 & 8x & 37.7$\pm$3.7 & 97.2$\pm$1.5 & 37.7$\pm$4.0 & 94.2$\pm$5.2 & 36.5$\pm$3.6 & 92.7$\pm$4.5 & 37.3 & 94.7\\
\hline
\multirow{2}{*}{FedDDA} & 4x & 37.5$\pm$5.2 & 95.3$\pm$2.5 & 39.9$\pm$3.6 & 95.4$\pm$1.3 & 39.1$\pm$3.2 & 95.1$\pm$1.8 & 38.8 & 95.3\\
 & 8x & 34.4$\pm$4.3 & 93.2$\pm$2.9 & 36.1$\pm$3.4 & 91.6$\pm$3.2 & 35.3$\pm$3.2 & 91.3$\pm$3.6 & 35.3 & 92.0\\
\hline
\multirow{2}{*}{FedGAT} & 4x & \textbf{45.5$\pm$4.2} & \textbf{99.1$\pm$0.7} & \textbf{43.5$\pm$4.8} & \textbf{98.1$\pm$1.7} & \textbf{43.4$\pm$4.2} & \textbf{98.0$\pm$1.3} & \textbf{44.1} & \textbf{98.4}\\
  & 8x & \textbf{38.8$\pm$3.6} & \textbf{97.5$\pm$1.5} & \textbf{38.8$\pm$4.3} & \textbf{94.6$\pm$5.1} & \textbf{37.9$\pm$4.0} & \textbf{93.8$\pm$4.1} & \textbf{38.5} & \textbf{95.3}\\
\hline
\end{tabular}}
\vspace{-0.25cm}
\end{table}

%% file: tables/table6.tex
\begin{table}[t]
\centering
\setlength\tabcolsep{6pt}
\renewcommand{\arraystretch}{1.4}
\caption{Susceptibility to information leakage for GAT priors trained with and without site prompts. FID scores of synthetic images were measured in the absence (w/o int) and presence (w int) of an intruder site in the FL setup. 
$\Delta$(w int $-$ w/o int) denotes intruder-induced degradation in image quality.
}

\resizebox{\columnwidth}{!}{%
\begin{tabular}{|c|c|c|c|c|c|c|c|}
\hline
\textbf{} & \multicolumn{2}{c|}{{Site 1}} & \multicolumn{2}{c|}{{Site 2}} & \multicolumn{2}{c|}{{Site 3}} & {Average} \\
\cline{2-8}
 & w/o int  & w int  & w/o int  & w int  & w/o int & w int  & $\Delta$\\
\hline
FedGAT & 0.58 & 3.51 & 0.46 & 5.02 & 0.03 & 1.40 & 2.95\\

w/o prompt & 2.16 & 5.98 & 0.33 & 9.71&3.91 & 21.12 & 10.14\\
\hline
\end{tabular}%
}
\label{tab:fid-delta}
\end{table}

%% file: tables/table7.tex
\begin{table}[t]
\centering
\caption{Reconstruction performance of FedGAT variants built to examine the contribution of individual method components.}
\resizebox{1.0\columnwidth}{!}{%
\begin{tabular}{| l l | c | c | c | c | c | c |}
 \hline
\multicolumn{2}{|c|}{\textbf{}} & \multicolumn{2}{c|}{Site 1 (fastMRI-knee)} & \multicolumn{2}{c|}{Site 2 (fastMRI-brain)} & \multicolumn{2}{c|}{Site 3 (UMRAM)} \\
\cline{3-8}
\multicolumn{2}{|c|}{\textbf{}} & PSNR & SSIM & PSNR & SSIM & PSNR & SSIM \\
\hline
\multirow{2}{*}{FedGAT} & 4x & \textbf{42.8$\pm$3.9} & \textbf{97.8$\pm$1.9} & \textbf{43.0$\pm$1.2} & \textbf{98.1$\pm$0.6} & \textbf{43.5$\pm$2.8} & \textbf{98.3$\pm$1.5} \\
& 8x & \textbf{37.2$\pm$2.3} & \textbf{94.0$\pm$4.9} & \textbf{38.5$\pm$0.9} & \textbf{95.2$\pm$1.5} & \textbf{38.8$\pm$2.9} & \textbf{95.7$\pm$3.9} \\
\hline
\multirow{1}{*}{w one-hot} & 4x & 41.3$\pm$3.2 & 97.5$\pm$1.9 & 42.1$\pm$0.9 & 97.7$\pm$0.6 & 40.7$\pm$1.9 & 97.4$\pm$2.1 \\
\multirow{1}{*}{prompt} & 8x & 36.2$\pm$2.7 & 93.3$\pm$5.3 & 36.9$\pm$0.8 & 94.4$\pm$1.8 & 35.7$\pm$1.3 & 94.7$\pm$4.1 \\
\hline
\multirow{1}{*}{w/o site} & 4x & 36.1$\pm$0.1 & 95.1$\pm$2.3 & 37.2$\pm$2.0 & 96.3$\pm$1.4 & 36.6$\pm$1.3 & 96.4$\pm$1.0 \\
\multirow{1}{*}{prompt} & 8x & 33.1$\pm$0.3 & 91.5$\pm$4.8 & 34.1$\pm$1.6 & 93.5$\pm$2.5 & 33.1$\pm$1.3 & 93.4$\pm$2.8 \\
\hline
\multirow{2}{*}{w/o scales} & 4x & 37.5$\pm$0.7 & 94.3$\pm$0.7 & 39.1$\pm$2.2 & 96.2$\pm$1.3 & 36.6$\pm$0.5 & 92.6$\pm$0.5 \\
& 8x & 33.4$\pm$0.3 & 91.2$\pm$3.4 & 35.4$\pm$1.5 & 92.9$\pm$2.9 & 33.9$\pm$0.6 & 91.2$\pm$2.9 \\
\hline
\multirow{2}{*}{w/o autoreg} & 4x & 39.0$\pm$1.0 & 96.3$\pm$1.1 & 41.7$\pm$1.6 & 97.6$\pm$1.0 & 39.0$\pm$1.2 & 94.8$\pm$0.2 \\
& 8x & 34.4$\pm$0.6 & 92.0$\pm$3.3 & 36.1$\pm$0.4 & 94.7$\pm$2.2 & 35.0$\pm$1.2 & 91.4$\pm$1.9 \\
\hline
\multirow{2}{*}{w model agg} & 4x & 39.3$\pm$1.7 & 96.9$\pm$1.7 & 39.3$\pm$1.7 & 97.2$\pm$0.4 & 40.5$\pm$3.4 & 97.8$\pm$2.1 \\
& 8x & 33.8$\pm$0.8 & 92.0$\pm$4.0 & 33.5$\pm$1.6 & 92.7$\pm$0.9 & 34.5$\pm$2.4 & 94.2$\pm$5.2 \\
\hline
\multirow{2}{*}{w synthetic} & 4x & 40.5$\pm$2.4 & 97.2$\pm$2.2 & 41.5$\pm$1.4 & 97.8$\pm$0.6 & 41.4$\pm$0.7 & 98.0$\pm$1.6 \\
& 8x & 35.7$\pm$2.2 & 93.2$\pm$5.3 & 36.4$\pm$0.0 & 94.9$\pm$1.5 & 36.6$\pm$1.5 & 95.4$\pm$3.8 \\
\hline
\end{tabular}
}
\label{tab:ablations}
\end{table}